\documentclass[a4paper,11pt]{article}

\usepackage{amsmath,amsfonts,amssymb,times,mathptmx,graphicx,amsthm}
\usepackage{placeins}
\usepackage[T1]{fontenc}
\usepackage[latin1]{inputenc}
\usepackage{lineno}
\usepackage{hyperref}

\title{\bf Unifying positioning corrections and random number generations in silicon micro-strip trackers }
\author{Gregorio Landi$^a$\thanks{Corresponding
author. Gregorio.Landi@unifi.it}~,   Giovanni E. Landi$^b$\\
\\
\llap{$^a$} Dipartimento di Fisica e Astronomia,
Universita' di Firenze \\
Largo E. Fermi 2 (Arcetri) 50125, Firenze, Italy\\
\\
\llap{$^b$} ArchonVR S.a.g.l.,\\
Via Cisieri 3,
6900 Lugano, Switzerland.\\ \\
{\em July 6, 2023}}
\date{ }
\begin{document}
\maketitle 
\begin{abstract}

The optimizations of the track fittings require complex simulations of silicon
strip detectors to be compliant with the fundamental properties of the hit
heteroscedasticity.
Many different generations of random numbers must be available with
distributions as similar as possible to the test-beam data.
A fast way to solve this problem is an extension of an algorithm of frequent use
for the center of gravity positioning corrections. Such extension gives a
single method to generate the required types of random numbers. Actually,
the starting algorithm is a random number generator, useful in a reverse mode:
from non uniform sets of data to uniform ones.
The inversion of this operation  produces random numbers of given distributions.
Many methods have been developed to generate random numbers, but none of
those methods is directly connected with this positioning corrections.
Hence, the adaptation of the correction algorithm to operate in both mode
is illustrated. A sample distribution is generated  and its consistency is
verified with the Kolmogorov-Smirnov test.
As final step, the elimination of the noise is explored, in fact, simulations
require noiseless distributions to be modified by given noise models.

\end{abstract}




\pagenumbering{arabic} \oddsidemargin 0cm  \evensidemargin 0cm

%
\section{Introduction }

The maximization of the resolution in track
reconstruction~\cite{landi15} requires
accurate eliminations of any systematic error present in the
observations (hit reconstructions).
Moreover, the inequalities of references~\cite{landi08,landi09}
impose careful accounts of the hit
variances (heteroscedasticity)
to reach the fit optimality. Evidently, the systematic errors disrupt
those essential improvements.
The easiest and frequently used positioning algorithm is the
so called center of gravity (COG); the weighted average of some strip
signals.
Unfortunately, this algorithm, as any other,
contains well known systematic errors~\cite{landi01},
whose dangerous effects on momentum reconstruction
are illustrated in
reference~\cite{landi06}.
However, such errors are disposable or reducible
to negligible effects, these corrections are
performed as described in~\cite{landi03}. The method
generalizes the original algorithm (the $\eta$-algorithm)
of reference~\cite{belau} removing its intrinsic
limitations.
The definition~\cite{landi03} of the $\eta$-algorithm
starts from a differential equation
that transforms the non uniform
COG distribution in an uniform one.
As expected, the COG algorithm
reconstructs the uniform distribution of hit-position
on the strips in a non uniform distribution,
clearly illustrated by the COG histogram.
This distortion is due to the detector noise
and the discretization of the signal distributions~\cite{landi01}.
The reverse transformation to an
uniform distribution compensate in large part the
effects of the discretization, leaving outside only the
noise effects that assumes the form of a narrow
zero-average distribution~\cite{landi03}.
Despite its flaws, the COG distortion in the histogram
contain approximate signatures of the noise effects on
the hit standard deviations. These signatures can be
easily inserted in track-fitting gaining a substantial
improvements of the track-parameters as demonstrated
in~\cite{landi15,landi16} (the lucky-model and
the super lucky-model).

Actually, real reconstructions are not so simple
as previously delineated. Instead, they require a set of
different COG algorithms, each one containing a
different number of strip signals in their calculations.
Those COG algorithms have very different analytic and
statistical properties~\cite{landi16} and must be tuned on the
specific case.  For simplicity,
we often speak of this complex set of COG algorithms
as a single element given that the mathematical procedure
for their corrections are practically identical.
However, we used the $\eta$-algorithm also well outside the
correction of the COG systematic error. Its property
allowed the extraction from the data of
average signal distributions released by a minimum
ionizing particle (MIP) on a set of nearby strips.
Such distributions are essential
to generate random distributions of correlated
signals among the strips, the foundation of our simulations.
In any case, the basic uses of the $\eta$-algorithm
are the correction of the hit positions of
incident MIPs on a layers of silicon strip
detectors, layers whose sizes
are many times the width of each
strip. This setup suggests naturally the periodicity
of the algorithm (with the period of a strip) and the
use of the Fourier series to handle the mathematics
of the problem.
In its essence, the algorithm
regenerates an uniform set of random numbers to an input of random
numbers with the COG distribution~\cite{landi03}. The periodicity
assures this correspondence for each strip of the layer. Apart
a set of mathematical details we added to the  $\eta$-algorithm
generalization, the algorithm remains very similar to that used in many
experiments. Those details allow to extend
the structure of the algorithm well beyond its original aim.
In particular, we are able to invert the algorithm and to obtain
a random number generator with the COG distribution. Our
interest was not addressed to generate random numbers, abundantly
produced by the data tacking, but to its derivative
that is connected to the average signal distribution
released by the MIP~\cite{landi03,landi05} in silicon detectors.
As previously mentioned, just this derivative allowed the extraction
from the data of another function very important for track
fitting: the average fraction of the signal distribution
collected by a cluster of strips, as a function of the
hit position of the MIP.  A theorem of reference~\cite{landi05}
is pivotal to this extraction. This function
implements also the two dimensional extension
to our probability distributions (as reported in~\cite{landi16}),
this extension  is the base for the maximum likelihood search in
track fitting~\cite{landi15,landi06,landi05}.
To complete its use, the fraction of the signal
distribution requires
another essential element: the effective
total signal released by the MIP and collected by the strips
of our interest (three strips very often)
to scale their fractions to the full value.
Hence, to complete the simulations for track
fitting it is required the
production of many distributions of signal released
by MIPs in double sided micro-strip detector~\cite{landi14,pamela}.
The theoretical forms of these distributions are
supposed well known, they are usually defined
as  Landau distributions convoluted with
Gaussian probability density functions (PDFs)
of the noise, and they are often fitted
in such a way.  In reality the distributions are
segments of  convolutions, the true convolutions
have infinite ranges.

Hence, given our need of only
a part of the hit cluster, quite often the strip
with the maximum signal and the two adjacent,
these PDFs differ from the theoretical ones. In addition,
for an easy consistency with the noise, we asked their
expressions in ADC counts as our data.
The cuts of the high energy side to reduce
$\delta$-rays give a finite range to the PDFs,
and the detection cuts eliminate the low energy
sides of the distributions. The available forms
of the PDFs are histograms with a
finite set of data, and this has to be the model
of our random number generator.
Evidently, this type of problem has a well known
solution: first produce the cumulative
distribution function and invert it. The last
step is not easy nor with a well defined method.
The appropriate method must be selected within
the large number of different solutions present
in literature, many of them can be found in
ref.~\cite{generators}.

Similar necessity, requiring the generation
of random numbers with a given
empirical PDF, is very frequent. Data analysis
are important sources of empirical PDF that
have to be used in the construction of realistic simulations.
Histograms with a finite number of events
(often not too many) are their usual forms.
Even if their fundamental probability
models are known, experimental details
modify the models in substantial way and
render the analytical models of doubtful use
in the production of realistic
data sets or too complex.

For the special physical properties of our
reconstructions, the developments of the
$\eta$-algorithm looks particularly efficient
for this class of problems. Although we illustrated
the method in references~\cite{landi01,landi03,landi05,landi04},
those illustrations are not effective to describe
the present use. There, the main task is
the hit position reconstructions, and this
obscures the extraction of the relevant steps.
Here, we concentrate on the procedure essential for
this application.
Section 2 describes the details of this unification,
in Section 3 the data generation is illustrated with the
intrinsic difficulties of our selected test distribution.
The Kolmogorov-Smirnov test of the generated distribution is in Section 4.
Two methods for noise filtering are discussed in Section 5.
Section 6 contains the summary and the conclusions.

\section{Definition of the problem}

\begin{figure}[h]
\begin{center}
\includegraphics[scale=0.6]{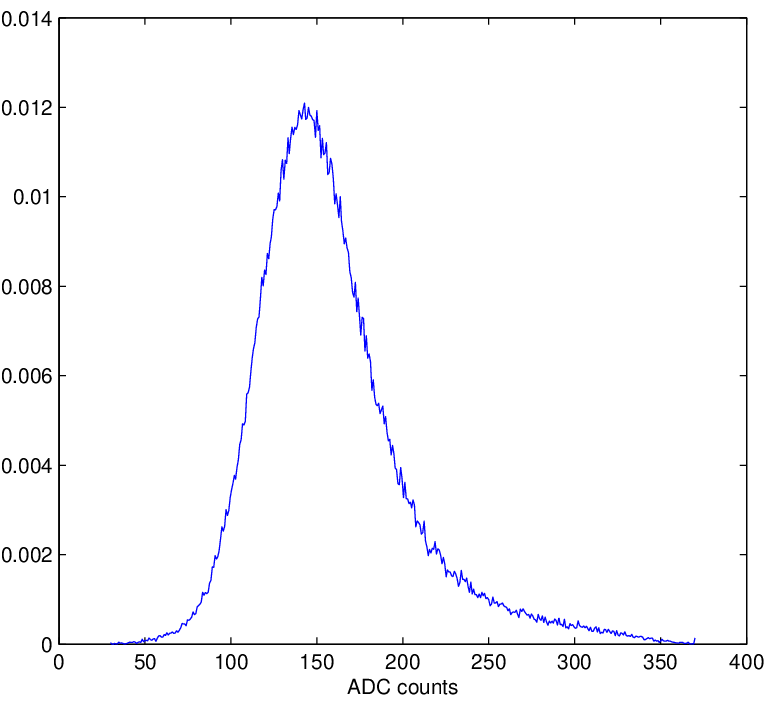}
\caption{ Empirical PDF for the signal released in three
strips by an incident MIP (ADC counts, 215613 events,
bin size 0.8 ADC counts)
}\label{fig:figure_1}
\end{center}
\end{figure}

A given PDF defines the distribution of
the generated random numbers,
the PDF must have finite range,
can be expressed in any scale and has to be
always different from zero.
Our model distribution is a set of
signals release by a MIP in two sided micro-strip
detectors. The data  were
acquired with a test-beam~\cite{pamela} in CERN
by a tracker with five detection layers. Our data cover
many incident directions and two different types of detectors.
For this illustration, we use the side with higher noise and
orthogonal incidence.
All the hits of the detector layers are collected together
to increase the statistics, and the charges detected
by three strips are reported on an histogram (normalized
and divided by the bin size) often indicated as frequency
polygon. The frequency polygon will be called
empirical PDF or simply PDF, and reported
in fig.~\ref{fig:figure_1}.

The data are distributed from 30 ADC counts to 370 ADC counts and the
PDF is different from zero in all the interval
as required. This type of interval is unpractical and it is better
to transform the interval in a standard one defined
from $-1/2$ and $+1/2$ with the trivial transformations $x=(x'-200)/340$
(and $y=y'*340$),
this interval coincides with that used for our COG corrections where
we normalize all the lengths with the strip width.
With this quite simple scaling of the PDF, the algorithm can be used
in any case. Due to these definitions, the dimensions in the
plots become pure numbers and do not require special indications.

Reference~\cite{landi01} generalizes the
$\eta$-algorithm of reference~\cite{belau}
extending it to a set of different COG algorithms
with any number of strips in the COG definition,
well beyond the two strips accounted  in~\cite{belau}, and releasing
its very complicated sorting of the strips.
The necessity of using COG algorithms with a
different number of strips is due to the substantial differences
in systematic errors and probability distributions (as in~\cite{landi16}
and therein references) of those
algorithms, they must be fine tuned on the applications.
The generalized $\eta$-algorithm eliminates the
systematic errors of the COG algorithms, but, for the use
at its best as in~\cite{landi16} (also with the insertion of an approximate
account of the hit heteroscedasticity to improve the fits),
the algorithm must be completed with some
details illustrated in references~\cite{landi01,landi03}.
Due to these generalizations, we have to handle events with
their COG positions outside the reference
interval $-1/2,+1/2$; the three strip COG
and the four strip COG have often such events.
The selected solution was the extension of the
supposed uniform illumination on a single virtual
strip to a supposed uniform illumination
on an infinite array of strips.
Thus, the data escaping from a strip are
recovered in the adjacent strips defining a
reproducible rule to handle uniformly in all these cases.
This simple modification introduces important
consequences on the method to handle the
COG probabilities: the periodicity.
In fact, to handle the hits on a large array of micro-strips,
the algorithm must be extended to work identically on
any strip cluster of an array with any number of micro-strips.
The periodicity eliminates one of the Kolmogorov axioms of
probability~\cite{gnedenko}: the normalization of the distribution.
We have to limit to a weaker condition
of normalization on a strip. We encountered the
necessity of this weaker condition also in
reference~\cite{landi16} again for the extension
of probabilities to detectors of any size.

\begin{figure}[h]
\begin{center}
\includegraphics[scale=0.6]{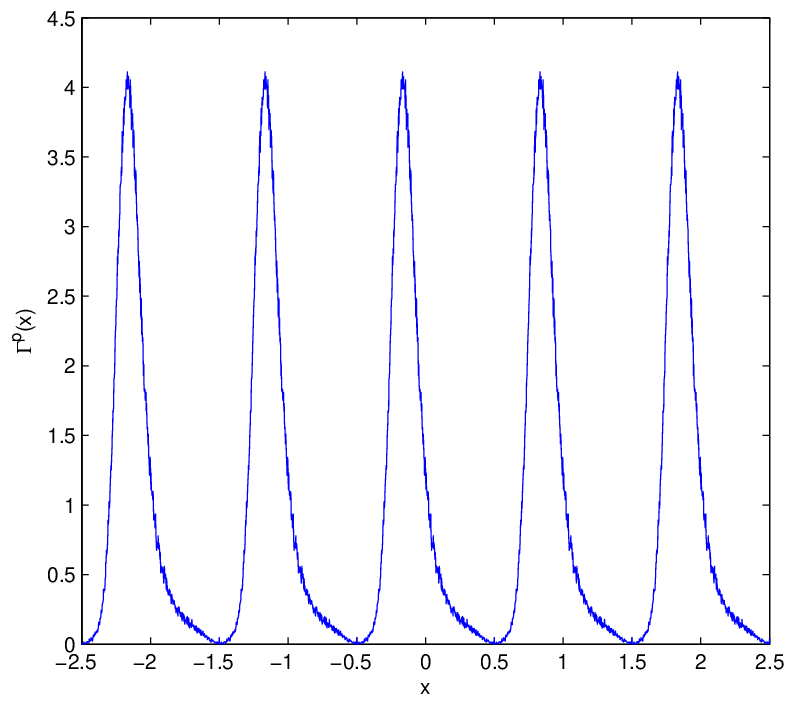}
\caption{ Periodic empirical PDF  in the transformed coordinate system ($\Gamma^p(x)$)
}\label{fig:figure_2}
\end{center}
\end{figure}

However, the fundamental gain of the periodicity is
the use of the Fourier Series to solve the equations
and the possibility to invert the COG equations
for any impact point.
These properties can be extended to the present problem as shown
in figure~\ref{fig:figure_2}.
Here, we have no underlying position calculation and we are
free from the complications due to the selection of the integration
constant of the differential equation. The selection of this
constant in positioning problem is easy  only in the case of
symmetrical set up, and this is an essential limit
of reference~\cite{belau}.

The change of variables in a PDF is given by the equation:
\begin{equation}\label{eq:equation_1}
    \Gamma(\varepsilon)\frac{\mathrm{d}\varepsilon}{\mathrm{d} x}=\Gamma^p(x),
\end{equation}
the probability  $\Gamma(\varepsilon)$ is transformed in a distribution
of $x$ with a probability  $\Gamma^p(x)$. The uniform
distribution is given by $\Gamma(\varepsilon)=1$, and the
function $\varepsilon(x)$ has a uniform distribution. The
equation~\ref{eq:equation_1} becomes the
first order differential equation:
\begin{equation}\label{eq:equation_1a}
    \frac{\mathrm{d}\varepsilon}{\mathrm{d} x}=\Gamma^p(x)
\end{equation}
with solution:
\begin{equation}\label{eq:equation_2}
    \varepsilon(x)=-\frac{1}{2}+\int_{-1/2}^{x}\,\Gamma^p(y) \mathrm{d}y\,.
\end{equation}
The integration starts at $-1/2$ to keep track of the single
period of our interest, and evidently contains
the generalized cumulative distribution of $\Gamma^p(x)$.
The assumption of
normalization on a strip allows to rewrite
equation~\ref{eq:equation_2} in the form:
\begin{equation}\label{eq:equation_3}
    \varepsilon(x)=x+\int_{-1/2}^{x}\,\big[\Gamma^p(y)-1\big] \mathrm{d}y
\end{equation}
It is easy to show that the integral in equation~\ref{eq:equation_3} is
periodic with period 1.
In fact, the contribution of all the complete periods,
contained in the interval $-1/2\vdash\dashv x$, are zero due
to the normalization of $\Gamma^p(x)$ on a period.
The integration on the remaining part of the interval is
identical in any period.
The Fourier Series becomes:
\begin{equation}\label{eq:equation_4}
\begin{aligned}
    &\varepsilon(x)=x+\sum_{k=-\infty}^{+\infty}\alpha_k
    \mathrm{e}^{2\pi \mathrm{i}\,k\,x} \ \ \ \ \ \ \ \ \alpha_k=\int_{-1/2}^{1/2}\big[\varepsilon(y)-y\big]
    \mathrm{e}^{-2\,\pi\, \mathrm{i}\,k\,y}\,\mathrm{d}y \ \\ &\alpha_k=\frac{1}{2\,\pi\,\mathrm{i}\,k}\int_{-1/2}^{1/2}\big[\Gamma^p(y)\big]\mathrm{e}^{-2\,\pi \,\mathrm{i}\,k\,y}\,\mathrm{d}y \ \ \ \ k\neq 0\ \ \ \ \ \ \ \  \alpha_0=-\int_{-1/2}^{1/2}\Gamma^p(y)\,y\,\mathrm{d}y
\end{aligned}
\end{equation}
The variable $\varepsilon$ coincides with $x$ at $-1/2$ and $+1/2$ and
at the end of any other period. The second line of the
equation~\ref{eq:equation_4} is obtained with an integration by part
of $\varepsilon(x)-x$ of equation~\ref{eq:equation_3}. The limitation on $k$
in numerical calculation operates as a filter that suppresses part of the
noise always present in empirical PDF.

The complex discussions of references.~\cite{landi03,landi04} were
specific for the position reconstructions where systematic errors
must be reduced at their minima. In that case the integration constant
of the differential equation must be an exact position (or the best
approximation) and it has to be extracted from the set of data
with unknown relations with the true impact points (excluding
the trivial case of symmetric signal distributions).


\section{The generator of random numbers}

Equation~\ref{eq:equation_4} clearly illustrates the gain of the
imposed periodicity, the function $\varepsilon(x)$
has a linear increasing component and a periodic part.
The sum of these two components has everywhere a positive
derivative (figure~\ref{fig:figure_2}).
Thus, $\varepsilon(x)$ can be inverted in another linear
increasing part and a periodic part:
\begin{equation}\label{eq:equation_5}
    x(\varepsilon)=\varepsilon+\sum_{m=-\infty}^{+\infty}\beta_m
    \mathrm{e}^{2\pi \mathrm{i} m \varepsilon}
    \ \ \ \ \ \beta_m=\int_{-1/2}^{+1/2} \big[x(\varepsilon)-
    \varepsilon\big] \mathrm{e}^{-2\pi \mathrm{i} m \varepsilon}\,
    \mathrm{d}\varepsilon
\end{equation}
In its form equation~\ref{eq:equation_5} is useless because we
do not know $x(\varepsilon)$, but we know $\varepsilon(x)$
and transforming the integration variable to $x$, the expression
for $\beta_m$ becomes:
\begin{equation}\label{eq:equation_6}
\begin{aligned}
    \beta_m=&\int_{-1/2}^{+1/2} \big[x-\varepsilon(x)\big] \mathrm{e}^{-2\pi \mathrm{i} m \varepsilon(x)}\,
    \frac{\mathrm{d}\varepsilon}{\mathrm{d} x}\mathrm{d}x\\
    =&\int_{-1/2}^{+1/2} \big[x-\varepsilon(x)\big] \mathrm{e}^{-2\pi \mathrm{i} m \varepsilon(x)}\,
    \Gamma^p(x)\mathrm{d}x\\
\end{aligned}
\end{equation}
where in the last line of equation~\ref{eq:equation_6} we introduced equation~\ref{eq:equation_1}.
The insertion of equation~\ref{eq:equation_3} gives:
\begin{equation}\label{eq:equation_7}
   \beta_m=-\int_{-1/2}^{+1/2}\mathrm{d}x \int_{-1/2}^{x}\,\mathrm{d}y\,\big[\Gamma^p(y)-1\big]  \mathrm{e}^{-2\pi \mathrm{i} m \varepsilon(x)}\,
    \Gamma^p(x)
\end{equation}
Other forms for $\beta_m$ can be obtained integrating by
parts equation~\ref{eq:equation_7} (the exponential and $\Gamma^p(x)$
are proportional to the $x$-derivative) and remembering
the boundary conditions $\varepsilon(\pm 1/2)=x(\pm 1/2)=\pm1/2$:
\begin{equation}\label{eq:equation_8}
    \beta_m=\frac{1}{2\pi \mathrm{i} m}\int_{-1/2}^{+1/2}\exp[-2\pi \mathrm{i}\, m\, \varepsilon(x)] \mathrm{d}x  \ \ \ \ \ \ m\neq 0 \ \ \ \ \ \ \beta_0=\int_{-1/2}^{+1/2}x\,\, \Gamma^p(x) \mathrm{d} x
\end{equation}
Evidently, the amplitude $\beta_0$ coincides with the first momentum of the PDF
as is clear in equation~\ref{eq:equation_5}. In fact, for $m=0$ the
integral of $\varepsilon$ is zero being an odd function, thus the sole
integration of $x\,\mathrm{d}\varepsilon/\mathrm{d}x$ remains and
the use of equation~\ref{eq:equation_1a} gives the result of
equation~\ref{eq:equation_8}.

These different expressions are useful in real calculations. The
unlimited sums in the Fourier series with the different forms of $\beta_n$ give
identical results, but the finite approximations
can show different sensitivity to the
details of the PDF.
We will call $\pm L_\alpha$ and $\pm L_\beta$
the extremes of the sums of equation~\ref{eq:equation_4}
and equation~\ref{eq:equation_5}. These finite values tend
to introduce very small shift of the boundary conditions
used in equation~\ref{eq:equation_8}, but due to periodicity these
are irrelevant or easily recovered.

\subsection{Numerical calculations}

The integration of equation~\ref{eq:equation_2} is  numerical, and
performed on a histogram with very small bin-size
(much smaller than that of figure~\ref{fig:figure_2}). Due to the
non decreasing property of the integral, the small bin-size is
able to follow accurately the form of $\Gamma^p(x)$.
Identical histogram is used also in  equation~\ref{eq:equation_4}
to calculate $\alpha_n$ with the limits $\pm L_\alpha$.
From now on $\varepsilon(x)$ will be given by its Fourier
series. Identical steps, but with COG histograms, allow to obtain
expressions to be used in~\cite{landi15,landi16}
for the lucky-model and the super lucky-model;
in fact, the derivatives
of similar Fourier series give analytical forms of
$\Gamma^p(x)$ reproducing the COG histograms
for the use in improving the track fitting.

The value of $L_\alpha$ operates as
an ideal low-pass filter that suppresses the fluctuations
always present in empirical PDF and those due to small bin
size of the numerical integration. Symmetrically, a too low
value of $L_\alpha$ can eliminate important
details of the PDF, hence, some care can be exerted on
this selection.

The inverse function $x(\varepsilon)$ is obtained with
the $\beta_m$ given by equation~\ref{eq:equation_7},
and the limits $\pm L_\beta$. The use of a low value of $L_\beta$ generates
random numbers with a set of  $L_\beta$ narrow
peaks. The peaks decrease in amplitude at increasing $L_\beta$ and
are smeared by the randomness of the $\varepsilon$
values, but they remain visible. These fluctuations
are typical of Fourier Series when the convergence of
the series is slow~\cite{libroFS}. The slow convergence is due
to large values of the transformed function, essentially
the inverse of the distribution of figure~\ref{fig:figure_2}.
The values around $\pm 1/2$ are very small and their
inverses become very large.
To suppress almost completely these fluctuations,
the large $L_\beta$ must be reinforced summing the
Fourier Series with the method of arithmetic mean
( Cesaro summation ) that is able to suppress the Gibbs
effect around discontinuities~\cite{libroFS}.
The remaining oscillation (if any) are invisible
also enlarging
at the maximum the following figures. 
In its finite form the Cesaro summation is equivalent to
the triangular filter $(1-|m|/L_\beta)$
(Bartlet window in digital filters)
applied to the $\beta_m$-amplitudes and centered at $m=0$.
The arithmetic mean redefines the form of the sequence of
functions converging to the Dirac $\delta$-function for
$N\rightarrow\infty$.
The standard sum produces functions of the form
$F_N(u)=\sin[(N+1/2)u/2]/\sin(u/2)$.
At any $N$, these functions have large positive and
negative oscillations around the peak in $u=0$ with
$1/u$ as weak smearing. The Cesaro summation
produces the set of converging functions to the
Dirac $\delta$-function of the form
$C_N(u)=[\sin(Nu/2)^2]/[2N\sin(u/2)^2]$.
Now the oscillations are strongly damped
by $1/u^2$ and are always positive.  A
a draw back is the slower increase of the peak,
in fact, $F_N(0)=N+1/2$ and $C_N(0)=N/2$. These
different types of convergence are relevant for finite sums.

\begin{figure}[h!]
\begin{center}
\includegraphics[scale=0.5]{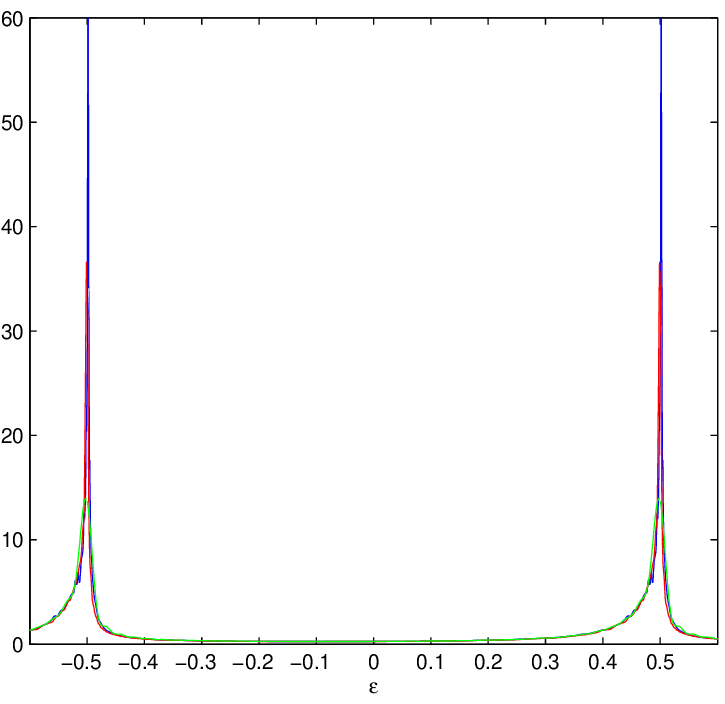}
\caption{ The blue line is ${1}/{\Gamma^p(x(\varepsilon))}$ the inverse of original PDF with its large values at $\pm1/2$, the other two curves are $ \mathrm{d} x(\varepsilon)/\mathrm{d}\varepsilon$ obtained with $L_\beta=200$ (the red line) and with
$L_\beta=40$ (the green line).
}\label{fig:figure_3}
\end{center}
\end{figure}

The complexity of the $\beta_m$ calculation is not
visible in equation~\ref{eq:equation_8}, but can be clearly
evidenced by the following expression:
\begin{equation}\label{eq:equation_9}
    \beta_m=\frac{1}{2\pi \mathrm{i} m}\int_{-1/2}^{+1/2}\exp[-2\pi \mathrm{i}\, m\, \varepsilon(x)] \frac{\mathrm{d}x}{\mathrm{d}\varepsilon} \mathrm{d}\varepsilon \ \ \ \ \
    \ \ \ \ \frac{\mathrm{d}x}{\mathrm{d}\varepsilon}=\frac{1}{\Gamma^p(x(\varepsilon))}
\end{equation}
The function ${1}/{\Gamma^p(x(\varepsilon))}$ is reported
in figure~\ref{fig:figure_3}, the other two curves (all
almost coinciding but with different heights at $\pm 1/2$)
are the numerical derivatives of $x(\varepsilon)$ obtained
from equation~\ref{eq:equation_8} with $L_\beta=200$ and $L_\beta=40$.
All these curves are  functions of $\varepsilon$.
The origin of the two peaks is due to the small
values of ${\Gamma^p(x)}$ at $x\approx \pm 1/2$.
To produce an uniform distribution, the cumulative distribution
$\varepsilon(x)$ is forced to compress the $x$-regions
of low probability in a narrow $\varepsilon$-region
enhancing the steepness of the peaks. For their good
reproduction, $L_\beta$ must be around 200 or higher.
The low values of ${\Gamma^p}$ towards $\pm1/2$
render this case more complex than our previous use of
this inversion for the COG problems. In micro-strip
detectors, the COG probabilities have restricted regions
of low probabilities generating negligible disturbances.

\begin{figure}[h!]
\begin{center}
\includegraphics[scale=0.6]{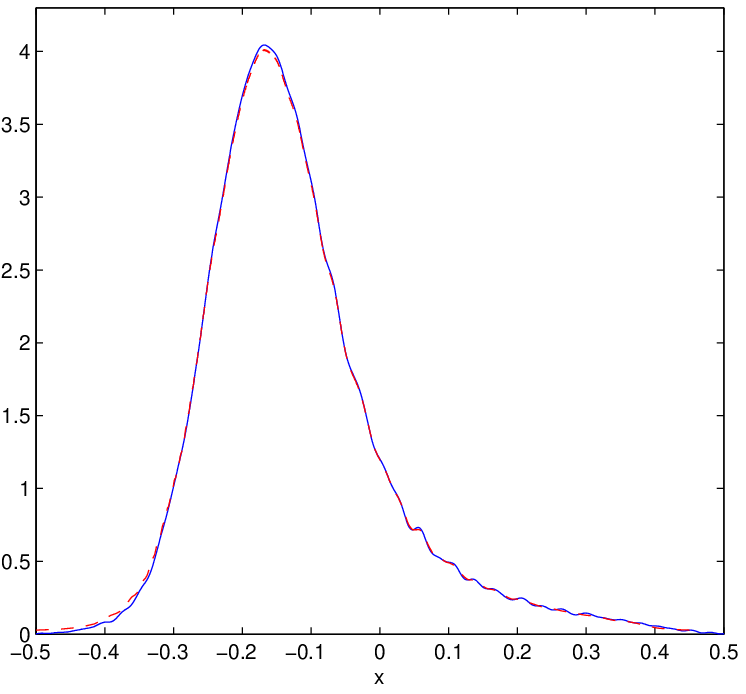}
\caption{ The blue line is $\Gamma^p(x)$ as given by the left
side of equation~\ref{eq:equation_1a} (essentially the data plot with a filter
on the main fluctuations), the red dashed line is the
expression $(\mathrm{d}x(\varepsilon)/\mathrm{d}\varepsilon)^{-1}$
plotted as a function of $x(\varepsilon)$ with $L_\beta=200$.
}\label{fig:figure_4}
\end{center}
\end{figure}

An interesting comparison is  with the
$\Gamma^p(x)$ obtained by the $\mathrm{d}\varepsilon(x)/\mathrm{d}x$
as in equation~\ref{eq:equation_1a} and the
$(\mathrm{d}x(\varepsilon)/\mathrm{d}\varepsilon)^{-1}$
plotted as function of $x(\varepsilon)$,
here $x(\varepsilon)$ is calculated with $L_\beta=200$.
The second function is the form of the data PDF given by this
random data generator. Figure~\ref{fig:figure_4}
illustrates the excellent agreement of the two curves.
This is not surprising given that $x(\varepsilon)$ is the inverse
function of $\varepsilon(x)$ thus $\varepsilon(x(\varepsilon))=\varepsilon$.
The root mean square of the difference
$\varepsilon(x(\varepsilon))-\varepsilon$ on a period is $1.7\,10^{-3}$.

\begin{figure}[h!]
\begin{center}
\includegraphics[scale=0.6]{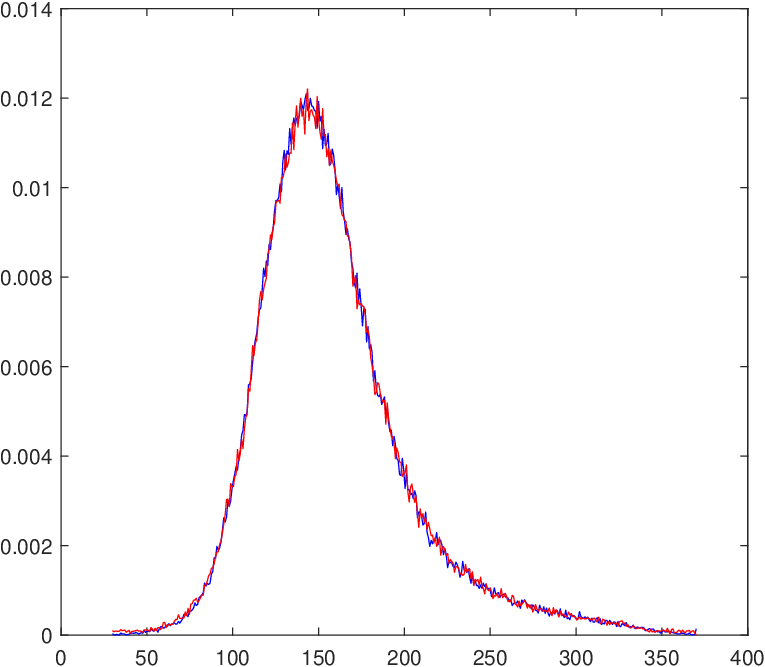}
\caption{ Histogram of the original data (the blue line) and the
histogram of the generated data (the red line) reported to
the right scale. The two histograms have identical number of
data, normalized and divided by the bin size (0.8 ADC counts).
}\label{fig:figure_5}
\end{center}
\end{figure}

As final test, figure~\ref{fig:figure_5} reports the histograms
(frequency polygons) of figure~\ref{fig:figure_1} and the
similar histogram with the data produced by this
generator and reported to the original scale.
The overlap is complete.

\subsection{PDF with long low probability tails}

We showed the difficulty introduced by these relatively low
tails of the PDF. Longer tails with lower probability
will be dangerous for this method. We can get around
this problem sectioning the tails and treating them as independent
PDF to be handled as described above. In this case, the data generation
proceeds in two steps. At the first, the interval 0-1 is
divided in two parts (or more if required) a main interval equal
to the cumulative  PDF of the main part, and a remaining interval
equal to the cumulative PDF of the tails. The extraction
of a uniform random number in the interval 0-1 selects which
PDF generators must be used~(Russian roulette).

\section{Kolmogorov test  }

Let us test the quality of our cumulative distribution function
$\varepsilon(x)$ respect to the empirical cumulative distribution
function obtained by the data. We have to rearrange the
data to obtain an ordered statistics $\{o_1,o_2\cdots o_N\}$
sorting the data with increasing values. This reorder renders
the empirical cumulative distribution quite simple: $F(o_k)=k/N$.
A Kolmogorov theorem~\cite{kolmogorov,mathstat}  states that for
$N\rightarrow\infty$ the probability of $\sqrt{N}\sup_{i}|F(o_i)-\varepsilon_1(o_i)|\leq t$
converges to $k(t)$. Where $\varepsilon_1(x)=\varepsilon(x)+1/2$
to recover the initialization of $\varepsilon(-1/2)=-1/2$,
and to report $\varepsilon_1(x)$ spanning the interval $0\vdash\dashv 1$.
More precisely, the probability $\mathbf{P}(x)$ is:
\begin{equation}\label{eq:equation_11}
    \lim_{N\rightarrow\infty}\mathbf{P}(\sqrt{N}\sup_{i}|F(o_i)-\varepsilon_1(o_i)|\leq t)=
    k(t)=1-2\sum_{j=1}^{+\infty}(-1)^{j-1} \exp(-2j^2t^2)
\end{equation}
For our $N$-values well above 200000 we can use the asymptotic
results and the maximum of our differences, scaled by $\sqrt{N}$,
is 0.383. For $t=0.383$ it is $k(0.383)\approx0.0015$, or
a probability around $0.15 \% $ to find a better cumulative function.

With small modification of Smirmov~\cite{mathstat} (a factor $1/\sqrt{2}$),
equation~\ref{eq:equation_11}  can be applied also to our generated set
of random data of figure~\ref{fig:figure_5}, but with a negative result.
The test is too sensible to the almost invisible differences of the low
probability parts of the distributions. We illustrated this problem in
figure~\ref{fig:figure_3} where the two high peaks are not well
reproduced. The differences are limited to a small number of events,
invisible in the plots, but able to move the test
toward the rejection region. To generate distributions fully
accepted by the Kolmogorov-Smirnov test, we have to proceed as in subsection
3.2, i.e. divide the probability distribution in three parts.
A central part with almost all the events, and two lateral side
with relatively few events. Applying the described algorithm at
the central part and the two lateral parts, assembled together to
produce a single PDF, it is easy to produce accepted distributions. The differences among
the plots reported in figure~\ref{fig:figure_5} and in
the following figure~\ref{fig:figure_8} are invisible to the eye,
\begin{figure}[h!]
\begin{center}
\includegraphics[scale=0.6]{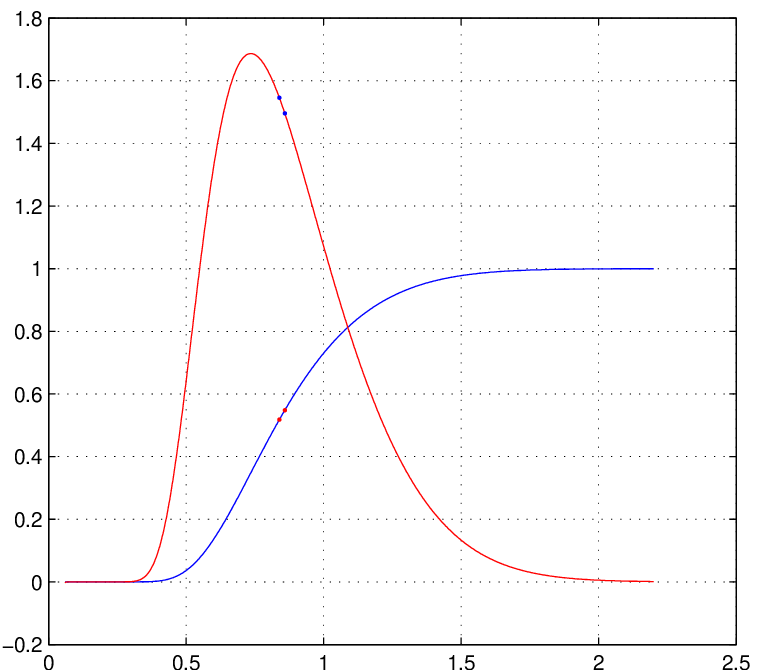}
\caption{ Plots of the Kolmogorov  cumulative distribution
probability of the asymptotic difference of maximum $\mathbf{P}(x)$  (blue line)
and the PDF of $\mathbf{P}(x)$  (red line). The two dots
are the values of $\mathbf{P}(x)$ for two different generations of random data.
One of the two is reported in figure~\ref{fig:figure_8}.
}\label{fig:figure_6}
\end{center}
\end{figure}
\begin{figure}[h!]
\begin{center}
\includegraphics[scale=0.6]{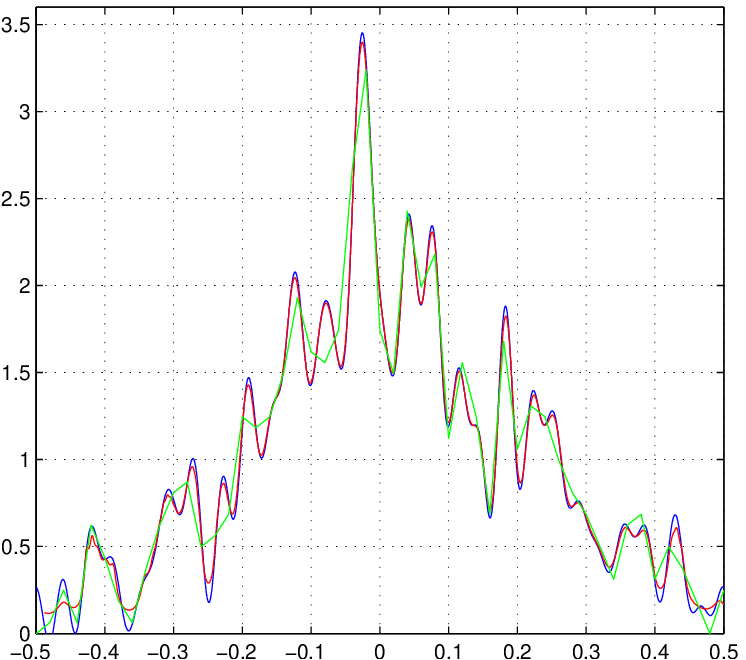}
\caption{ Plots of the low probability distributions, blue line
$(d\varepsilon/dx)$, red line the reconstructed distribution
$(d\,x(\varepsilon)/d\varepsilon)^{-1}$ the green line is the histogram of the data
}\label{fig:figure_7}
\end{center}
\end{figure}
\begin{figure}[h!]
\begin{center}
\includegraphics[scale=0.6]{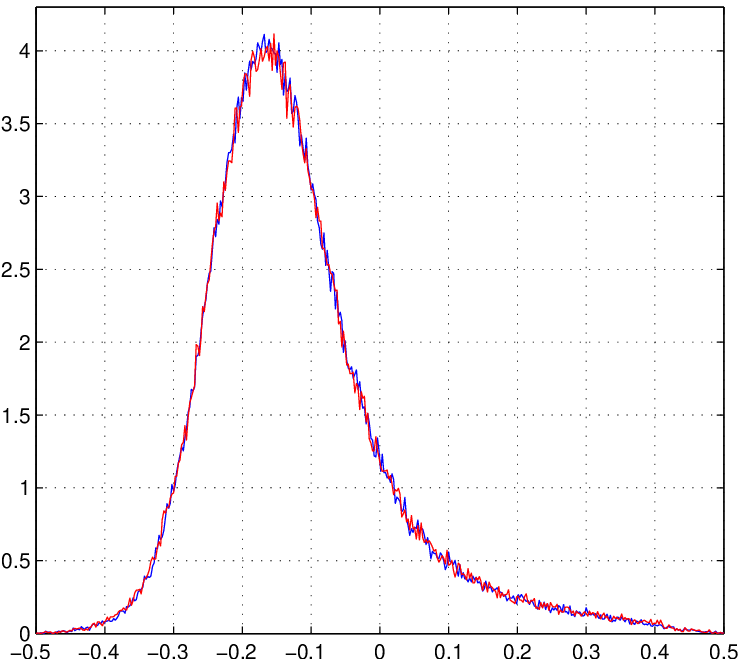}
\caption{ Plots of the full probability distributions (low+high) ,
blue line is the histogram of the data. The red line is the
histogram of the generated random numbers
accepted by the Kolmogorov test
}\label{fig:figure_8}
\end{center}
\end{figure}

The separation of the low probability data and the high
probability data, produces sets of random numbers easily accepted by the
Kolmogorov-Smirnov test. Equation~\ref{eq:equation_11}, for a value of
the argument 0.859 of our example generation, gives a probability of
0.548 well below the 0.95, boundary of the confidence level, and 0.859 is among the
most probable values of these differences as shown in figure~\ref{fig:figure_6}.
Figure~\ref{fig:figure_7} is the plot of the low probability parts and
figure~\ref{fig:figure_8} illustrates the comparison of the histogram of the
data with the generated random numbers accepted by the test.

\section{Noise suppression }

\subsection{Lucy-Richardson method}

The analytical form given by the equation~\ref{eq:equation_4}
rises the possibility to extract the noiseless form from
the noisy distribution of figure~\ref{fig:figure_1}.
The large set of data reduces strongly the noise fluctuation of
the histogram of figure~\ref{fig:figure_1}, although noise components
survive. A variation of the method of Lucy and
Richardson~\cite{Lucy,Richard} will be use to reduce those components.

In principle, the noise effect is a convolution of the noiseless
signal with the Gaussian noise distribution. We tested the noise
distribution in our data exploring the random signals of the strips outside
the signal clusters. The probability of another hit in these strips
is evidently negligible and the random signals are essentially noise.
These random signals are well reproduced by a
Gaussian PDF. The Fourier transform of a convolution gives an apparent
strategy to clean the noise, but further details must be consider.
The first one is the finite range of our data set. For an analytical
function, even of finite range, the convolution with a Gaussian gives
a function different from zero on all the real line,
with negligible values on the tails. These tails are essential to give
the correct convergence at high frequency so sustaining the division
by another Gaussian (the Fourier transform of a
Gaussian noise). Instead, a finite range distribution has
a Fourier transform that goes to $\infty$ as $1/\omega$. Multiplication
by $\exp(\omega^2)$ generates an explosion of the high frequency and
a very oscillating result. With some care, the Lucy-Richardson method
allows an analytical solution for our case.
\begin{equation}\label{eq:equation_10}
    \Gamma^p(x)=\Pi(x)\int_{-1/2}^{+1/2}P(x-y)\Psi(y) \mathrm{d}y
\end{equation}
This equation is essentially that of reference~\cite{Lucy},
$\Gamma^p(x)$ is the signal corrupted by the noise, and $P(x-y)$ is the
Gaussian noise PDF with variance $\sigma^2$, $\Psi(y)$ is
the noiseless distribution and $\Pi(x)$ is the interval function
($\Pi(x)=1$ for $|x|\leq 1/2$, $\Pi(x)=0$ for $|x|<1/2$).
In addition to the Gaussian noise, $\Gamma^p(x)$ contains the noise
due to the finiteness of the data sample. This last noise is
difficult to be modeled and must be dealt empirically in a
more or less reasonable methods.

\begin{figure}[h!]
\begin{center}
\includegraphics[scale=0.5]{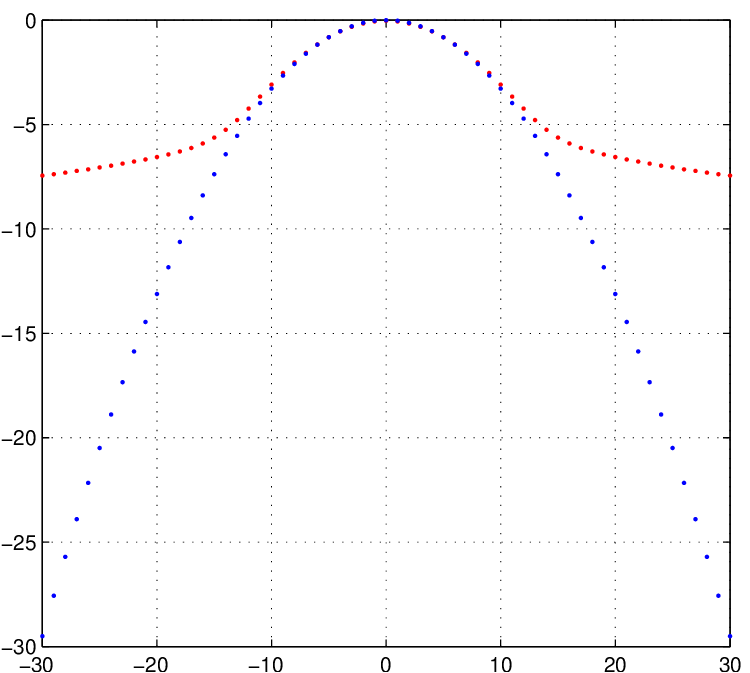}
\caption{ The red dots are the $log(N_{k,k})$ for the
noise of $8\sqrt{3}$ ADC counts of the sum of three strip signals. The blue dots are the
the logarithms of the corresponding pure Gaussian factors.
}\label{fig:figure_8a}
\end{center}
\end{figure}

\begin{figure}[h!]
\begin{center}
\includegraphics[scale=0.6]{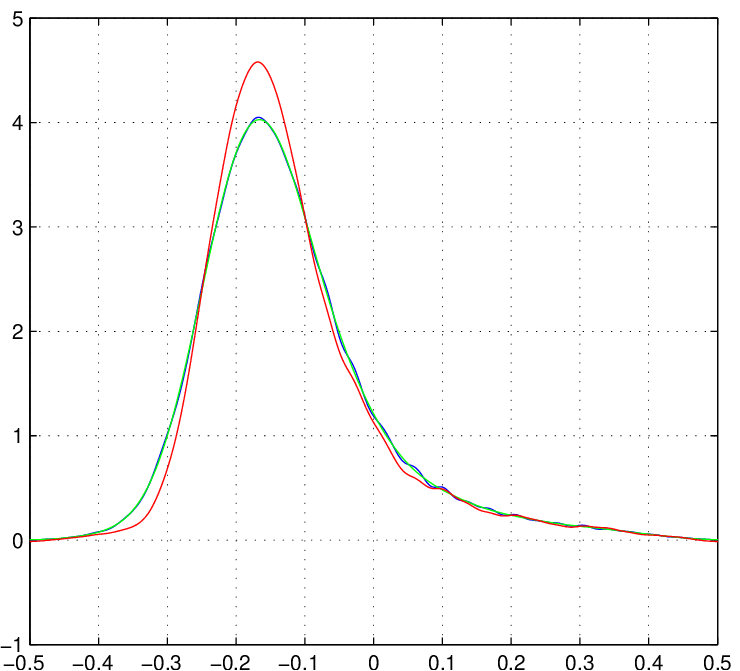}
\caption{ The red line is a possible noiseless result,
the blue line is the original PDF, and the green line is given by the full
reapplication of the noise to $\gamma_n^0$  at all the amplitudes.
}\label{fig:figure_9}
\end{center}
\end{figure}

Consistently with our assumption on $\Gamma^p(x)$, we will
assume an identical periodicity also for $\Psi(y)$. Inserting
the Fourier Series for $\Gamma^p(x)$ (amplitudes
$\gamma_n=2\pi \mathrm{i} n\alpha_n$ and $\gamma_0=1 $ ) and
$\Psi(y)$ (with amplitudes $\gamma^0_n$),
equation~\ref{eq:equation_10} becomes:
\begin{equation}\label{eq:equation_11a}
    \gamma_n=\sum_k N_{n,k}\gamma_k^0 \ \ \ \ \
    N_{n,k}=\int_{-1/2}^{+1/2}\exp[-2\pi \mathrm{i}( n x- k y)]P(x-y) \mathrm{d}x \mathrm{d} y
\end{equation}
The integrals of equation~\ref{eq:equation_11a} are not too complex,
and MATHEMATICA~\cite{MATHEMATICA} is able to do them.
The off-diagonal terms are
irrelevant and  the diagonal terms alone are worth of attention,
they have the expression:
\begin{equation}\label{eq:equation_12}
\begin{aligned}
    N_{k,k}=&\sqrt{\frac{2}{\pi}}\,\sigma \big[\mathrm{e}^{-1/2\sigma^2}\cos(2\,k\pi) -1)\big]+
    \frac{1}{2}\mathrm{e}^{-2(k\pi\sigma)^2}\Big\{4k\pi\sigma^2 \mathrm{erfi}(\sqrt{2}k\pi\sigma)+\\
    &2 \mathrm{real}\big[(1-2 \mathrm{i} k\pi\sigma^2) \mathrm{erf}(\frac{1-2\mathrm{i}k\pi\sigma^2}{\sqrt{2}\sigma})\big]\Big\}\\
\end{aligned}
\end{equation}
As $k$ increases $N_{k,k}$ has an almost $1/k$ decrease (as expected), very
different from the rapid decrease of a Gaussian term as shown in
figure~\ref{fig:figure_8a}. For $k\leq 11$, $N_{k,k}$ coincides
with the pure Gaussian correction.
In spite of this, the direct use of $N_{k,k}$ in equation~\ref{eq:equation_10} is
very unrealistic. The noise introduced by the histogram
fluctuations are similar to a white noise with frequency
spectrum that further attenuates the decrease of $\gamma_k$ with $k$.
The large number of events assures a reliability of the first
$\gamma_k$ amplitudes, on which we can apply equation~\ref{eq:equation_10}.
Neglecting the frequencies with $k$ greater than $9$ or $10$, the
strip noise suppression gives the result of figure~\ref{fig:figure_9}.

The plausibility of the solution can be tested approximating
the distributions as a Gaussian PDFs (at least around the maxima),
in this case the noise suppression is an easy procedure
$\sigma_t^2={(\sigma_0^2+\sigma_n^2)}$
and the full width at half maximum of the noisy PDF is
$\sigma_t\sqrt{8\ln(2)}=0.211$ giving a  $\sigma_t=0.0896$.
Due to the data strip-noise $\sigma_n=8\sqrt{3}/340$,
than the noiseless $\sigma_0$ would be 0.0798 and a noiseless
full width at half maximum of 0.188. The full width at half maximum of
the red line in figure~\ref{fig:figure_9}
is 0.180 that is very similar to the one for a Gaussian function.

\subsection{Wiener filter}

\begin{figure}[h!]
\begin{center}
\includegraphics[scale=0.6]{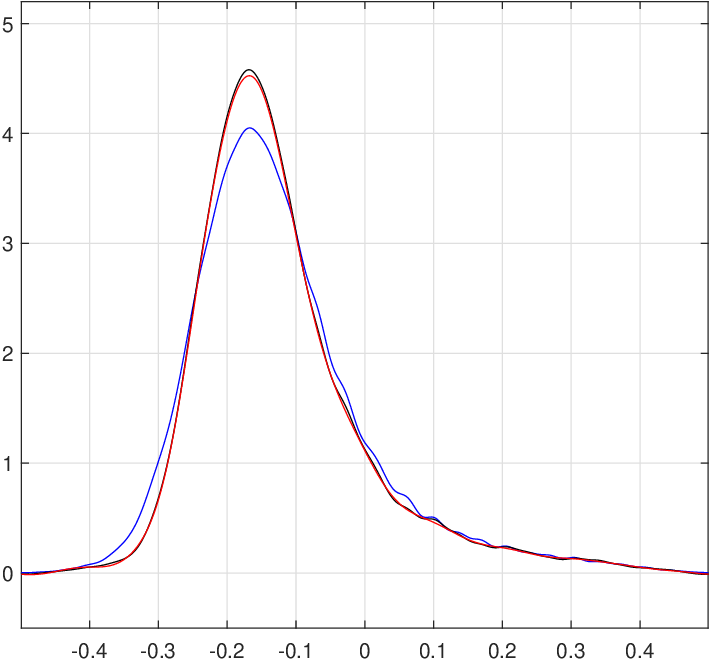}
\caption{ The red line is the result of the Wiener deconvolution,
the blue line is the original PDF, and the black line is the previous filtering.
}\label{fig:figure_10}
\end{center}
\end{figure}

Another interesting method to compensate the noise is the use of the
Wiener filter~\cite{wiener}. The Wiener filter executes a deconvolution
of the signal keeping into account also the additional noise due to the
signal elaboration (as the histogram construction). The insertion
of this additional noise moderates the effects of the division
by the rapidly decreasing amplitude of the Gaussian Fourier transforms.
As shown previously, also the expressions of the equation~\ref{eq:equation_12},
with their linear decrease, create rapidly strong oscillations with a
small number of k-terms. To have a reasonable result we have to add a drastic
limit to the k-maximum. The Wiener filter operates with the following equation
for the Fourier series amplitudes:
\begin{equation}\label{eq:equation_13}
    \gamma_k^0=\gamma_k\,\,\frac{\,(N_k^g)^*}{\,|N_k^g|^2\,+\,S_k}\,.
\end{equation}
The terms $N_k^g$ are the Gaussian amplitudes of figure~\ref{fig:figure_8a},
the parameters $S_k$ are the ratios of the mean power of the additional noise
(the detection/elaboration noise) to the mean power of the signal. We do
not invest effort to calculate the $S_k$ amplitudes and, as usual, we assume
a constant value for any $k$-term, as having to do with a white noise.
The $S_k=0.01$ gives a result very similar
to the previous development as illustrated in figure~\ref{fig:figure_10}.
The effect of the terms $S_k$ in equation~\ref{eq:equation_13} are
very clear, they suppress the effects of the small amplitudes
$N_k^g$. Equation~\ref{eq:equation_13} has the general form with the
complex numbers $N_k^g$. In our case, the $N_k^g$ are real numbers for
the symmetry of the Gaussian PDF.
The constant $S_k$ is not critical and the filtered
distribution has very few shifts for values around $0.01$. Instead
the value $0.001$ starts  to show substantial oscillations in
the filtered distribution.

The generations of the "noiseless" random signals now begin from
the distributions of figure~\ref{fig:figure_10} going back to
equation~\ref{eq:equation_6}. The filtered $\alpha_n^0$
is given by $\alpha_n^0=\gamma_n^0/(2\,\pi\,\mathrm{i}\,n)$
composing the series of $\varepsilon^0(x)$.

\subsection{Empirical noise "suppression"}

An empirical method can be used to reduce the noise effects, we
used this in reference~\cite{landi05}. The histogram of figure~\ref{fig:figure_1} is shrinked by the three strip
noise supposing the core of the histogram to be a
Gaussian PDF. The amplitudes were those reported at the
end of subsection 5.1. The shinking operates also on the
total interval of the data, reducing it to 303 ADC counts.
The new histogram is used to construct the random number
generator from equation~\ref{eq:equation_3}.
The random numbers,
simulating the strip noise, are generated starting from
equation~\ref{eq:equation_3} and a histogram of strip signals
for strips far from those interested by the MIP signals
(exclusively noise with high probability). This selection accounts
also any deviation from a pure Gaussian noise, for example
the truncation noise due to the ADC conversion,
and has finite range. The simulated MIPs data
produce histograms of single strip signals
with excellent overlaps on those of
the real MIP data.

\section{Conclusions}

The account of the hit heteroscedasticity introduces a complex
interplay among  the center of gravity as positioning algorithm,
its histograms, the signals collected by the strips and the hit
variances to be inserted in the track fitting.
To test and isolate those relations,  many sets
of simulations are required. The standardization of the methods
for random number generations strongly alleviated this heavy
work and could allow to isolate other useful relations.
The unification in a single algorithm the correction of the systematic error
of center of gravity algorithm and the generation of other necessary
random numbers is an excellent step in that direction.
The algorithm for the correction of the center of gravity
positioning was carefully discussed in previous publications,
but some specific aspects of this correction obscures its
usefulness for the present task.
The periodicity, imposed by layers of micro-strips, suggests
the use of the Fourier series and its powerful shortcuts.
The oscillations produced by the small values of the sample
of probability distributions can be cured in various ways,
increasing the numbers of Fourier amplitudes, using the Cesaro
summation or splitting the probability distribution in sections.
This last strategy is able to generate sets of random numbers
accepted by the Kolmogorov-Smirnov test, although
these distributions show no appreciable differences
with generations obtained without that splitting. Evidently,
the low probability section is strongly modified
by the small number of events and their large fluctuations.
The test is very sensitive to those
fluctuations, but in practical use the sectioning can reasonably
neglected. As final development, the filtering of the noise in the
original distribution is explored with two different methods,
the Lucy-Richardson method and the Wiener method with similar
results. The plausibility of the noise filtering is verified
approximating the cores of the distributions as Gaussian
distributions. In this case the noise suppression
is an easy analytical task, and it shows a clear consistency with
the more elaborated methods.





{\bf Abbreviations}\\

{ The following abbreviations are used in this article}:\\

\noindent
\begin{tabular}{@{}ll}
ADC & Analog-to-Digital Converter\\
COG & Center of Gravity\\
MIP & Minimum Ionizing Particle\\
PDF & Probability Density Function\\
\end{tabular}

\end{document}